\documentclass{elsart} 
\usepackage{graphicx} 
\usepackage{amssymb}

\def\lsim{\ \rlap{\raise 3pt \hbox{$<$}}{\lower 3pt \hbox{$\sim$}}\ }
\def\gsim{\ \rlap{\raise 3pt \hbox{$>$}}{\lower 3pt \hbox{$\sim$}}\ }

\def\plb#1{Phys.\ Lett.\ {B\,\bf #1}} 
\def\prc#1{Phys.\ Rev.\ {C\,\bf #1}} 
\def\prd#1{Phys.\ Rev.\ {D\,\bf #1}} 
\def\prl#1{Phys.\ Rev.\ Lett. {\bf #1}}

\def\rmp#1{Rev.\ Mod.\ Phys.\ {\bf #1}}

\def\sjnp#1{Sov. J. Nucl. Phys. {\bf #1}} 

\def\lsim{\raise0.3ex\hbox{$\;<$\kern-0.75em\raise-1.1ex\hbox{$\sim\;$}}}
\def\gsim{\raise0.3ex\hbox{$\;>$\kern-0.75em\raise-1.1ex\hbox{$\sim\;$}}}

\def\be7{$^{7}$Be}
\def\b8{$^{8}$B}
\begin{document}
\begin{frontmatter}

\title{Large mixing angle solution to the solar neutrino problem and
random matter density perturbations}
 
\author[unicamp]{M.M. Guzzo} \author[unicamp]{P.C. de Holanda}
\author[puc]{N. Reggiani}

\address[unicamp]{{\it Instituto de F\'\i sica Gleb Wataghin} \\ {\it
Universidade Estadual de Campinas - UNICAMP} \\ {\it 13083-970
Campinas, S\~ao Paulo, Brasil} } \address[puc]{{\it Centro de
Ci\^encias Exatas, Ambientais e de Tecnologias} \\ {\it Pontif\'\i cia
Universidade Cat\'olica de Campinas} \\ {\it Caixa Postal 317} \\ {\it
13020-970 Campinas, SP, Brasil}}

\begin{abstract}
There are reasons to believe that mechanisms exist in the solar
interior which lead to random density perturbations in the resonant
region of the Large Mixing Angle solution to the solar neutrino
problem.  We find that, in the presence of these density
perturbations, the best fit point in the ($\sin^2 2\theta, \Delta
m^2$) parameter space moves to smaller values, compared with the
values obtained for the standard LMA solution. Combining solar data
with KamLAND results, we find a new compatibility region, which we
call VERY-LOW LMA, where $\sin^2 2\theta \approx 0.6$ and $\Delta
m^2\approx 2\times 10^{-5}$~eV$^2$, for random density fluctuations of
order $5\% < \xi< 8\%$. We argue that such values of density
fluctuations are still allowed by helioseismological observations at
small scales of order 10~-~1000~km deep inside the solar core.
\end{abstract}

\end{frontmatter}

Assuming CPT invariance, the electronic antineutrino $\bar{\nu}_e$
disappearance as well as the neutrino energy spectrum observed in
KamLAND~\cite{KamLAND} are compatible with the predictions based on
the Large Mixing Angle (LMA) realization of the MSW mechanism,
resonantly enhanced oscillations in matter~\cite{Wolf,MS}. This
compatibility makes LMA the best solution to the solar neutrino
anomaly~\cite{HS}-\cite{SNO}.  The best fit values of the relevant
neutrino parameters which generate such a solution are $\Delta
m^2=7.3\times 10^{-5}~$eV$^2$ and $\tan^2\theta=0.41$ with a free
boron neutrino flux $f_B=1.05$~\cite{Holanda-Smirnov(02)} with the
1~$\sigma$ intervals: $\Delta m^2=(6.2 \leftrightarrow 8.4) \times
10^{-5}~$eV$^2$ and $\tan^2\theta=0.33 \leftrightarrow 0.54$.

KamLAND results exclude not only other MSW solutions, such as the
small mixing angle realization, but also the just-so solution and the
alternative solutions~\cite{30years} to the solar neutrino problem
based on non-standard neutrino interactions~\cite{GMP}-\cite{BGHKN},
resonant spin-flavor precession in solar magnetic
field~\cite{schechtervalle}-\cite{GN98} and violation of the
equivalence principle~\cite{6GFN}-\cite{vepmsw}. Therefore, the LMA
MSW solution appears, for the first time in the solar neutrino problem
history, as the unique candidate to explain the anomaly.

Such an agreement of the LMA MSW predictions with the solar neutrino
data is achieved assuming the standard approximately exponentially
decaying solar matter distribution~\cite{JNB}-\cite{BP00}.  This
prediction to the matter distribution inside the sun is very robust
since it is in good agreement with helioseismology
observations~\cite{helioseismology}.

When doing the fitting of the MSW predictions to the solar neutrino
data, it is generally assumed that solar matter do not have any kind
of perturbations. I.e., it is assumed that the matter density
monotonically decays from the center to the surface of the Sun.  There
are reasons to believe, nevertheless, that the solar matter density
fluctuates around an equilibrium profile. Indeed, in the
hydro-dynamical approximation, density perturbations can be induced by
temperature fluctuations due to convection of matter
between layers with different temperatures.  Considering a Boltzman
distribution for the matter density, these density fluctuations are
found to be around 5\%~\cite{Hiroshi}. 
Another estimation of the level of density
perturbations in the solar interior can be given considering the
continuity equation up to first order in density and velocity
perturbations and the p-modes
observations. This analysis leads to a value
of density fluctuation around 0.3\%~\cite{Turck-Chieze(93)}.

The mechanism that might produce such density fluctuations can also be
associated with
modes excited by turbulent stress in the convective
zone~\cite{Bahcall} or by a resonance between g-modes and magnetic
Alfv\`en waves~\cite{Valle}.
As the g-modes occur
within the solar radiative zone, these resonance creates spikes at
specific radii within the Sun. It is not expected that these
resonances alter the helioseismic analyses
because as they occur deep inside the Sun, they do not affect
substantially the observed p-modes.  
This resonance depends on
the density profile and on the solar magnetic field, and as mentioned in
Ref.~\cite{Valle}, for a magnetic field of order of 10 kG the spacing
between the spikes is around 100 km. 
In the analyses presented in Ref.~\cite{Valle} the values considered
for the magnetic field are the ones that satisfy the Chandrasekar
limit, which states that the magnetic field energy must be less than
the gravitational binding energy.

Considering helioseismology, there are constraints on the density
fluctuations, but only those which vary over very long scales, much
greater than 1000 km~\cite{Castellani,Bahcall1000,JCD}.  In
particular, the measured spectrum of helioseismic waves is largely
insensitive to the existence of density variations whose wavelength is
short enough - on scales close to 100 km, deep inside within the solar
core - to be of interest for neutrino oscillation, and the amplitude
of these perturbations could be large as 10\%~\cite{Valle}.

So, there is no reason to exclude density perturbations at a few
percent level and there are theoretical indications that they really
exist.

In the present paper, we study the effect on the Large Mixing Angle
parameters when the density matter fluctuates around the equilibrium
profile.  We consider the case in which these fluctuations are given
by a random noise added to an average value.  This is a reasonable
case, considering that in the lower frequency part of the Fourier
spectrum, the p-modes resembles that of noise~\cite{Hiroshi}.  Also,
considering the resonance of g-modes with Alfv\'en waves, the
superposition of several different modes results in a series of
relatively sharp spikes in the radial density profile. The neutrino
passing through these spikes fell them as a noisy perturbation whose
correlation length is the spacing between the density
spikes~\cite{Valle}.

Noise fluctuations have been considered in several cases. Loretti and
Balantekin~\cite{Loretti} have analyzed the effect of a noisy magnetic
field and a noisy density.  In particular, for a noisy density, they
found that the MSW transition is suppressed for a Small Mixing
Angle. For the case of strongly adiabatic MSW transitions and large
fluctuations, the averaged transition probability saturates at
one-half. Refs.~\cite{Hiroshi} and~\cite{Carlos}
have analyzed the effect of a matter density noise on the MSW effect
and found that the presence of noisy matter fluctuations weakens the
MSW mechanism, thus reducing the resonant conversion
probabilities. These papers, nevertheless, did not take into
consideration KamLAND data.

In order to analyze the effect of a noisy density on the neutrino
observations, we must consider the evolution equations for the
neutrino when the density is given by a main average profile perturbed
by a random noisy fluctuation. This is done starting from the standard
Schr\"odinger equation~\cite{Loretti,Hiroshi}.

The evolution for the $\nu_e$ - $\nu_y$ system is governed by

\begin{eqnarray}
& 
i{\frac{\displaystyle{\ d }}{\displaystyle{\ dr}}}
\left(
\begin{array}{c}
\nu_e \\ \nu_y
\end{array}
\right) = 
& 
\left(
\begin{array}{cc}
H_e & H_{ey} \\ H_{ey} & H_y
\end{array}
\right) \left(
\begin{array}{c}
\nu_e \\ \nu_y
\end{array}
\right),
\label{motion}
\end{eqnarray}
where

\begin{eqnarray}
H_e = 2[A_{ey}(t) + \delta A_{ey}], \\ H_y = 0, ~ H_{ey} \approx
{\frac{\Delta m^2}{4E}} \sin 2 \theta, \\ A_{ey}(t) =
{\frac{1}{2}}\left[V_{ey}(t) - {\frac{\Delta m^2}{4E}}\cos 2\theta
\right], \\ \delta A_{ey}(t) \approx \frac{1}{2} V_{ey}(t)\xi .
\end{eqnarray}

Here $E$ is the neutrino energy, $\theta$ is the neutrino mixing angle
in vacuum, $\Delta m^2$ is the neutrino squared mass difference and
the matter potential for active-active neutrino conversion reads

\begin{equation}
V_{ey}(t) = {\frac{\sqrt{2} G_F}{m_p}} \rho(t)(1-Y_n) ,
\end{equation}
where $V_{ey}$ is the potential, $G_F$ is the Fermi constant, $\rho$
is the matter density, $m_p$ is the nucleon mass and $Y_n$ is the
neutron number per nucleon. 
$\xi$ is the fractional perturbation of the matter potential.

To calculate the survival probability of the neutrinos it is necessary
to average the terms of the evolution equation over the random density
distribution.  Here we assume a delta-correlated Gaussian noise used
in Refs.~\cite{Loretti,Hiroshi}:

\begin{equation}
<\delta A_{ey}^{2n+1}> ~ =0, \\ ~ <\delta A_{ey}(t) \delta
A_{ey}(t_1)> ~ = 2k(t)\delta(t-t_1) ,
\end{equation}
where the quantity $k(t)$ is defined by
\begin{equation}
k(t) = ~ <\delta A_{ey}^2(t)> L(t) = {\frac{1}{2}} V_{ey}^2(t) <\xi^2>
L(t) ,
\end{equation}
where $L(t)$ is the correlation length of the perturbation. 
The averages, denoted by $<>$, are realized in space for one correlation
length, and the delta-correlated noise means that 
the fluctuations in density are completely 
spatially uncorrelated for separations larger than one 
correlation length.
$L(t)$
must obey the following relations

\begin{equation}
l_{free} << L(t) << \lambda_m(t),
\label{condition}
\end{equation}
where $l_{free}$ is the mean free path of the electrons in the Sun and
$\lambda_m(t)$ is the characteristic neutrino matter oscillation
length.

In order to guarantee that this condition is satisfied in the whole
trajectory of the neutrino inside the sun, we assume, following
Nunokawa et al.~\cite{Hiroshi}, that

\begin{equation}
L(t) = 0.1 ~\lambda_m(t) = 0.1~{\frac{2 \pi}{\omega(t)}},
\end{equation}
where $\omega(t)$ is the frequency of the MSW effect, given by

\begin{equation}
\omega^2(t) = 4(A^2_{ey}(t) + H^2_{ey}).
\end{equation}
Eqs. (10) and (11) give a value of $L(t)$ between 10 and 100 km in the
resonant region for the LMA effect, that is, for the inner part of the
Sun.

\begin{figure}[t] \centering
    \includegraphics[width=0.85\textwidth,scale=1.,angle=0]{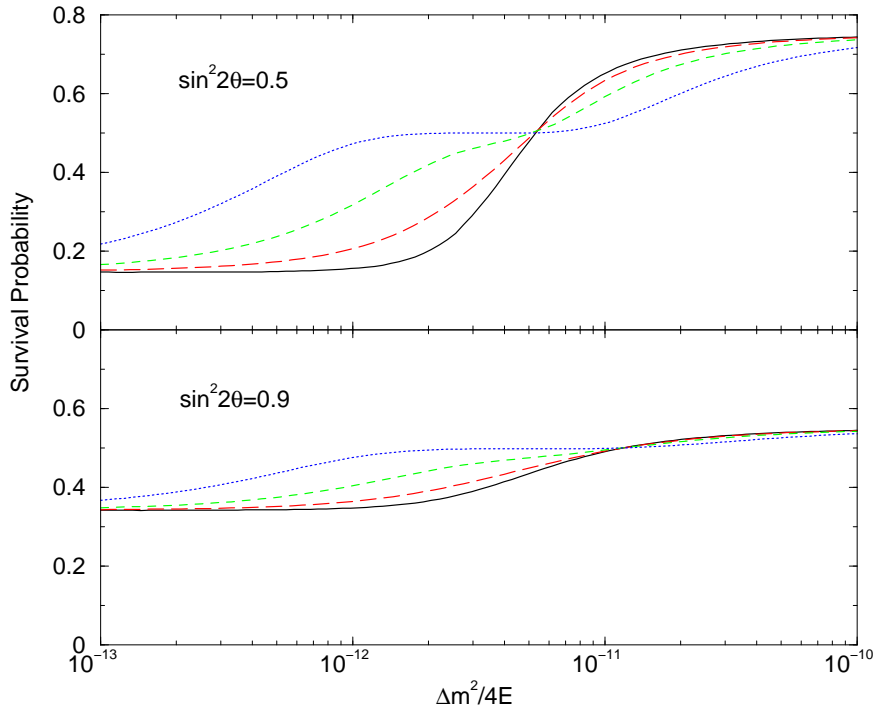}
    \caption{\label{fig:psurv} 
      values of the mixing angle, and several values of the
      perturbation amplitude, $\xi=0\%$ (solid line), $\xi=2\%$ (long
      dashed line), $\xi=4\%$ (dashed line) and $\xi=8\%$ (dotted
      line).}
\end{figure}

We calculate the survival probability of the neutrinos solving the
evolution equation, considering the equilibrium density profile given
by the Standard Solar Model~\cite{BP00}.  The effect of the density
noise can be seen in Fig.~1, where the survival probability of the
neutrinos is plotted as a function of $\Delta m^2/{4E}$ for four
values of density perturbation, $\xi = 0\%, 2\%, 4\% $ and $ 8\% $. We
can observe that the effect is amplified when $sin^2 2 \theta = 0.5$,
rather than when it is $sin^2 2\theta = 0.9$. Considering that the
best fit value of $\Delta m^2$ obtained from a combined analysis of
the solar neutrino data and KamLAND observations is of order
$10^{-5}$~eV$^2$, we can say that the neutrinos with energy around 10
MeV will be the most affected by the density noise.

\begin{figure}[t] \centering
    \includegraphics[width=0.85\textwidth]{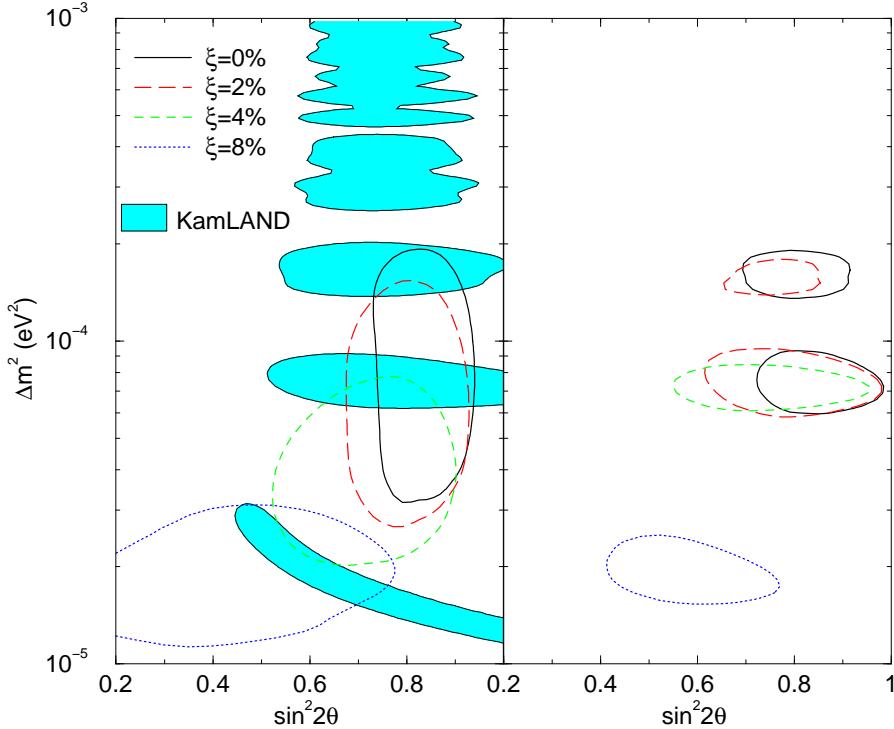}
    \caption{\label{fig:chi2} LMA region for different values of the
      perturbation amplitude, at 95\% C.L. for several values of the
      perturbation amplitude, $\xi=0\%$ (solid line), for $\xi=2\%$
      (long dashed line), $\xi=4\%$ (dashed line) and $\xi=8\%$
      (dotted line). We also present the allowed region for KamLAND
      spectral data, for the same C.L..  In the right-handed side of
      the figure, the combined analysis of both solar neutrino and
      KamLAND observations is shown. }
\end{figure}

In Fig.~2 we present the ($\sin^2 2\theta, \Delta m^2 $) parameter
space comparing the results obtained for solar neutrinos with the
allowed regions obtained from KamLAND observations, for four values of
density perturbation, $\xi = 0\%, 2\%, 4\% $ and $8\%$. We observe
that the values of the parameters $\Delta m^2$ and $\sin^2 2\theta$
that satisfy both the solar neutrinos and KamLAND observations are
shifted in the direction of lower values of $\Delta m^2$ and $\sin^2
2\theta$ as the amplitude of the density noise increases.  In the
left-handed side of Fig.~2, the best fit point of the solar analysis
with no perturbations lies in $(\sin^2 2\theta,\Delta m^2) =
(0.82,7.2\times 10^{-5} {\rm eV}^2)$, with a minimum
$\chi^2=65.4$. For $\xi=4\%$, the best fit point goes to $(\sin^2
2\theta,\Delta m^2) = (0.71,3.5\times 10^{-5} {\rm eV}^2)$, with a
minimum $\chi^2=64.0$, while for $\xi=8\%$ we have $(\sin^2
2\theta,\Delta m^2) = (0.33,1.4\times 10^{-5} {\rm eV}^2)$, with a
minimum $\chi^2=64.2$. The numbers of degrees of freedom (d.o.f.) in
this analysis is 78, obtained from 81 data points, 2 oscillation
parameters and $\xi$.

The right-handed side of Fig.~2 shows the combined analysis involving
both solar neutrino and KamLAND data.  The best fit point of this
analysis when no perturbations is assumed lies in $(\sin^2
2\theta,\Delta m^2) = (0.85,7.2\times 10^{-5} {\rm eV}^2)$, with a
minimum $\chi^2=71.1$. For $\xi=4\%$, our best fit point goes to
$(\sin^2 2\theta,\Delta m^2) = (0.81,7.2\times 10^{-5} {\rm eV}^2)$,
with a minimum $\chi^2=71.5$, while for $\xi=8\%$ we have $(\sin^2
2\theta,\Delta m^2) = (0.52,1.9\times 10^{-5} {\rm eV}^2)$, with a
minimum $\chi^2=75.5$. Here, the number of d.o.f. is 91 .

The main consequence of introducing random perturbation in the solar
matter density is the appearance of entirely new regions in the
($\sin^22\theta \times \Delta m^2$) which allow simultaneous
compatibility of solar neutrino data and KamLAND observations. Besides
the two standard LMA regions, shown in the right-handed side of Fig.~2
by the continuous lines, which were called high and low-LMA in
Ref.~\cite{Holanda-Smirnov(02)}, we find a new region displaced toward
smaller values of $\sin^22\theta$ and $\Delta m^2$ which we call
VERY-low-LMA. There $\Delta m^2 = (1 \leftrightarrow 3)\times 10^{-5}$
eV$^2$ and $\sin^22\theta =(0.4 \leftrightarrow 0.8)$, at 95\% C.L.,
obtained for $\xi=8\%$.

With no perturbation, the solar data restricts the value of the mixing
angle, while the KamLAND data can restrict the value of $\Delta
m^2$. A combined analysis leads to $\Delta m^2=(5\leftrightarrow
9)\times 10^{-5}$ eV$^2$ and $sin^2 2\theta = (0.67 \leftrightarrow
0.98)$ at the low-LMA region, and $\Delta m^2=(1 \leftrightarrow
2)\times 10^{-4}$ eV$^2$ and $\sin^2 2\theta = (0.70 \leftrightarrow
0.95)$ at the high-LMA region, at $95\%$ C.L..

Including the effects of random perturbations, the solar data is not
able anymore to restrict the value of the mixing angle, and both
mixing and $\Delta m^2$ are more restricted by the KamLAND data.  And,
since the LMA regions are displaced to lower values of mixing and
$\Delta m^2$, the allowed KamLAND region with lower values of $\Delta
m^2$ becomes important.
In fact, such a displacement has already been noticed by Nunokawa et
al. \cite{Hiroshi} without considering the now available data like as
zenith spectrum from solar neutrino observations as well as data from
KamLAND. Therefore their focus was not only on LMA solution but also
small mixing angle parameters with active and sterile neutrinos.

\begin{figure}[t] \centering
    \includegraphics[width=0.85\textwidth]{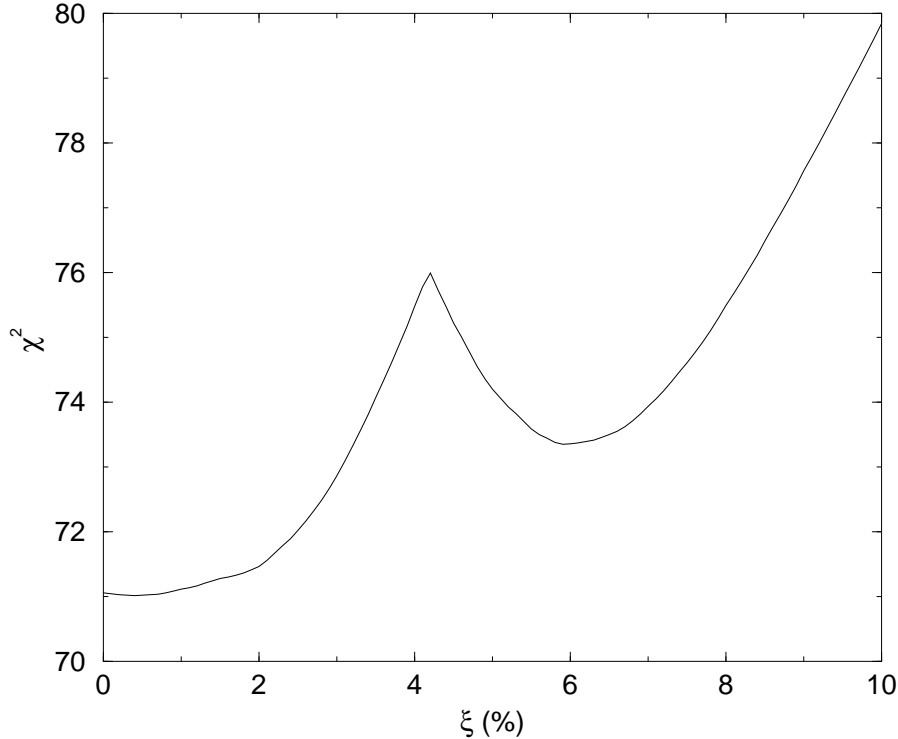}
    \caption{\label{fig:xsi}  $\chi^2$ as a function of $\xi$, the
       perturbation amplitude, where we minimize in neutrino
       parameters. Here the number of degrees of freedom is 91.}
\end{figure}

In Fig.~3 it is shown $\chi^2$ as a function of the perturbation
amplitude, minimized in $\Delta m^2$ and $sin^2 2\theta$. We can see
that even for high values of the perturbation amplitude we still can
have a viable solution. We notice that even in a noisy scenario the
compatibility of solar neutrino and KamLAND results is still good. In
fact, although the absolute best fit of the analysis lies on the
non-noise picture where $\xi = 0$, we observe that
$(\chi^2-\chi^2_{min}) < 4$ for $5\%<\xi <8\%$, showing a new scenario
of compatibility.

We conclude the paper arguing that random perturbations of the solar
matter density will affect the determination of the best fit in the
LMA region of the MSW mechanism.  Taking into account KamLAND results,
the allowed regions of $\Delta m^2$ and $sin^2 2\theta$ at 95\%
C.L. moves from two distinct regions, the low-LMA and high-LMA, when
no noise is assume in the corresponding resonant region, to an
entirely new VERY-low-LMA, when a noise amplitude of $5\%<\xi<8\%$ is
assumed inside the Sun.

This represents a challenge for the near future confront of solar
neutrino data and high-statistic KamLAND observations. If KamLAND will
determine $\sin^22\theta<0.5$ and $\Delta m^2 < 4\times
10^{-5}$~eV${^2}$ it can be necessary to invoke random perturbations
in the Sun to recover compatibility with solar neutrino observations.

\vspace{0.5cm}

{\bf Note added}: After completion of this paper we received two new
articles which analyzed similar pictures as ours. One is the
Ref.~\cite{Valle-new} which, differently from its first
version~\cite{Valle}, included KamLAND data.  In this paper a
step-function correlated noise has been considered which allows to
relax the condition given by Eq.~(\ref{condition}). Note that for the
LMA parameters, the correlation length obtained from
Eq.~(\ref{condition}) does not differ very much from the one used in
this reference which, therefore, arrive to similar numerical results
as ours, summarized in Fig.~3.

Also, a new paper~\cite{Balantekin-new} arrived to different
conclusions, concerning the quality of the fit for solar neutrino data
together with KamLAND data in a noisy scenario.

\vspace{0.5cm}

{\bf Acknowlegments}: The authors would like to thank FAPESP and CNPq
for several financial supports.

\end{document}